\documentstyle[12pt]{article}
\begin{document}
\bibliographystyle{unsrt}

\title{Effects of the quantity $\sigma_{TS}$ on the
spin structure functions of nucleons in the resonance region}
\author{\normalsize
Zhenping Li$^{[1]}$ and Dong Yu-Bing$^{[2]}$\\
\small{$^{1}$Physics Department, Carnegie-Mellon University},\\
\small{Pittsburgh, PA, 15213-3890}\\
\small{$^{2}$CCAST(World Laboratory)P.O.Box 8730, 
Beijing 100080,P. R. China}\\
\small{and}\\
\small{Institute of High Energy Physics, Academia Sinica,}\\
\small{ Beijing 100039, P. R. China\thanks{Mailing address}}}
\maketitle 
\par
\vspace{0.2cm}
\noindent
\begin{abstract}
In this paper, we investigate the effects of the quantity $\sigma_{TS}$ 
on the spin-structure functions of  nucleons  in the resonance region.
The Schwinger sum rule for the spin structure function 
$g_2(x,Q^2)$ at the real photon limit is derived for the nucleon treated 
as a composite  system, and it provides a crucial constraint on the 
longitudinal transition operator which has not been treated consistently 
in the literature.  The longitudinal amplitude $S_{\frac 12}$ is evaluated
in the quark model with the transition operator that generates the Schwinger
sum rule.   The numerical results of the quantity 
$\sigma_{TS}$ are presented for both spin structure functions $g_1(x,Q^2)$ 
and $g_2(x,Q^2)$ in the resonance region.  Our results show that this quantity
 plays an important role in the low $Q^2$ region,  which can be tested 
in the future experiments at CEBAF.
\end{abstract}


\newpage
\section*{1. Introduction}
 The quantity $\sigma_{TS}$, defined in the spin structure 
functions of nucleons
\begin{equation}\label{1}
g_{1}(x,Q^{2})=\frac{M_TK}{8\pi^{2}\alpha(1+\frac {Q^{2}}{\omega^{2}})}
\left [\sigma_{1/2}(\omega,Q^{2})-\sigma_{3/2}(\omega, 
Q^{2})+\frac{2\sqrt{Q^{2}}}{\omega}\sigma_{TS}(\omega,Q^{2})\right ]
\end{equation}
and 
\begin{equation}\label{2}
g_{2}(x,Q^{2})=\frac{M_TK}{8\pi^{2}\alpha(1+\frac {Q^{2}}{\omega^{2}})}
\left [\frac{2\omega}{\sqrt{Q^{2}}}\sigma_{TS}(\omega, Q^{2})-
(\sigma_{1/2}(\omega, Q^{2})-\sigma_{3/2}(\omega, Q^{2}))\right ],
\end{equation}
where K is the photon flux, $x$ the scaling variable, and $M_T$ the 
nucleon mass,  was not fully investigated due to the fact that most
studies were concentrated in the deep inelastic scattering region,
where the quantity $\frac{2\sqrt{Q^{2}}}{\omega}\sigma_{TS}(\omega,Q^{2})$
in $g_1(x,Q^2)$ vanishes. This is nolonger the case recently as there
 have been growing interests in studying the spin structure functions in 
the small $Q^2$ region, where the resonance contributions are important.  
Consequently, the investigation of the effects of the quantity $\sigma_{TS}$ 
 in the small $Q^2$ region has become increasingly important.

Such a program began with the suggestion\cite{an} that the $Q^2$ dependence 
of the spin dependent sum rule might play a significant role in the 
interpretation of the European Muon Collaboration(EMC) data\cite{emc}, which
starts with a negative Drell-Hearn-Gerasimov(DHG)\cite{dhg} 
sum  rule in the real photon limit and ends with a positive sum rule\cite{ej}
in the large $Q^2$limit\cite{zz93}:
\begin{equation}\label{111}
\int_0^1 g_1(x,Q^2)dx = \left \{ \begin{array} {r@{\quad}l}
-\frac {\omega_{th}}{4M_T}\kappa^2 & Q^2=0 \\
\Gamma & Q^2\to \infty\end{array} \right. 
\end{equation}
where 
\begin{equation}\label{5}
\omega_{th}=\frac{Q^{2}+2m_{\pi}M+m_{\pi}^{2}}{2M}
\end{equation}
is the threshold energy of pion photoproductions, $\kappa$ the 
anomalous magnetic moment, and $\Gamma$ a positive quantity.
Because the contributions from the quantity $\sigma_{TS}$ to the
sum rule in Eq. \ref{111} also vanish in the real photon limit, 
most quantitative studies\cite{zz93,zpli,burkert} of the $Q^2$ dependence 
of the sum rule in Eq. \ref{111} were concentrated on the contributions 
from the quantity $\sigma_{\frac 12}-\sigma_{\frac 32}$ in $g_{1,2}(x,Q^2)$.  
Indeed, these investigations have shown a strong $Q^2$ dependence of the 
sum rule in the $Q^2\le 2.5$ GeV$^2$ region.    However, the study by Soffer 
and  Teryaev\cite{sofer} suggested that the quantity $\sigma_{TS}$ plays a 
significant role in the small $Q^2$ region,  which is highlighted by another 
set of sum rules for the spin structure function $g_2(x,Q^2)$;   
\begin{equation}\label{112}
\int_0^1 g_2(x,Q^2)dx = \left \{ \begin{array} {r@{\quad}l}
\frac {\omega_{th}}{4M_T}\kappa (\kappa+e_T) & Q^2=0 \\
0& Q^2\to \infty , \end{array} \right. 
\end{equation}
in which the same kinematics in Eq. \ref{111} is used.  
The sum rules in Eq. \ref{112} were first derived by Schwinger\cite{schwinger}
 in the real photon limit and Burkhardt and Cummingham\cite{bc} in
 the large $Q^2$ limit.  Combining Eqs. \ref{111} and \ref{112} leads to the 
sum rule for the quantity $\sigma_{TS}$ in the real photon limit;
\begin{equation}\label{35}
\lim_{Q^2\to 0}\int_{\omega_{th}}^{\infty}\sigma_{TS}\frac{d\omega}{\sqrt{Q^2}}
=\frac {4\pi^2\alpha}{4M_T^2} e_T \kappa . 
\end{equation}
The magnitude of the sum rule for the quantity $\sigma_{TS}$ in the
 real photon limit is certainly comparable to the DHG sum rule.  Thus, 
a more quantitative study of the
contributions from the quantity $\sigma_{TS}$ to the spin structure 
functions $g_1(x,Q^2)$ and $g_2(x,Q^2)$ is called for.  Such a study 
not only enables us to give a more precise estimate of the $Q^2$ 
dependence of the sum rule for $g_1(x,Q^2)$, but also provides a
quantitative calculation of the sum rule for $g_2(x,Q^2)$ for the
 first time in the quark model.  The focus of this paper is to 
develop a framework in the quark model to evaluate the contributions
from the quantity $\sigma_{TS}$, and present the numerical results
that can be tested in the future CEBAF experiments.

The sum rules in Eqs. \ref{111} and \ref{112} are more general 
and model independent, therefore, they must be satisfied in the quark 
model in order to give a consistent evaluation of the spin structure 
functions in the resonance region.  This has been investigated\cite{zpli} 
in the quark model for sum rules in Eq. \ref{111},  and they require that 
the electromagnetic interaction for a many body system should has the form
\begin{eqnarray}\label{11}
H_t&=&\sum_{j}\{
e_{j}\vec{r}_{j}\cdot\vec{E}_{j}
-\frac{e_{j}}{2m_{j}}\vec{\sigma}_{j}\cdot\vec{B}_{j}-
\frac{e_{j}}{4m_{j}}\vec{\sigma}_{j}\cdot
[\vec{E}_{j}\times\frac{\vec{p}_{j}}{2m_{j}}-
\frac{\vec{p}_{j}}{2m_{j}}\times\vec{E}_{j}] \nonumber  \\
&&+\sum_{j<l}\frac{1}{4M_{T}}[\frac{\vec{\sigma}_{j}}{m_{j}}
-\frac{\vec{\sigma}_{l}}{m_{l}}]\cdot(e_{l}\vec{E}_{l}\times\vec{p}_{j}
-e_{j}\vec{E}_{j}\times\vec{p}_{l})\},
\end{eqnarray}
 and the quantity $\Gamma$ in Eq. \ref{111} is related to the quark model
matrix element
\begin{equation}\label{113}
\Gamma=\frac 12 \langle i| \sum_j e_j^2 \sigma_j^z|i \rangle_{P-A}
\end{equation}
where quark $j$ at position $r_j$ has mass $m_j$ and charge $e_j$, and
$A(P)$ indicates that the directions of the polarization between
photons and the target are antiparallel(parallel).  On the other hand,
the sum rule for the quantity $\sigma_{TS}$ in Eq. \ref{35} has not been 
investigated in the quark model. The derivation of Eq. \ref{35} in the 
quark model is by no means trivial since it  was proved\cite{schwinger}
 in QED by assuming the nucleon as  an elementary particle. The similar
 transition of the DHG sum rule from an elementary particle to a many 
body system led to 
extensive discussions in late sixties and early seventies\cite{brody}. 
Moreover, the proof of Eq. \ref{35} requires evaluations of both helicity 
amplitude $A_{\frac 12}$ and the longitudinal amplitude $S_{\frac 12}$.
While the helicity amplitude $A_{\frac 12}$ has been calculated\cite{cl91} 
with the transition operator $H_t$ in Eq. \ref{11}, the longitudinal 
amplitude $S_{\frac 12}$ has not been treated consistently in the literature.  
In particular,  the problem of the current conservation was  not fully
 understood\cite{muko},  and an {\it ad hoc} current  
$J^\prime_3=-\frac {k_3J_3-k_0J_0}{k_3}$ was  introduced\cite{abu} to 
evaluate the longitudinal transitions of the baryon resonances\cite{warns}.
The sum rule for the quantity $\sigma_{TS}$ provides an important test to
the quark model; the consistency requires that the sum rules for both 
$g_1(x,Q^2)$ and $g_2(x,Q^2)$ be generated by the same set of transition
operators for a many body system.  In section 2, we show that Eq. \ref{35}
is indeed generated by the electromagnetic interaction $H_t$ in Eq. \ref{11},
which also satisfies the DHG sum rule.
The longitudinal transition operator is obtained by requiring it satisfying 
the sum rule in Eq. \ref{35}, which is not only gauge invariant, but also 
consistent with $H_t$ in Eq. \ref{11}.  In particular, 
the spin-orbit interaction and the Wigner rotation that are crucial to 
the DHG sum rule for a many body system should be present in the longitudinal 
transition operator.

In section 3, we show that the quantity $\sigma_{TS}$ in $g_2(x,Q^2)$ cancels
 the transverse cross section $\sigma_{\frac 12}-\sigma_{\frac 32}$ in the
large $Q^2$ limit, which leads to the well known Burkhardt-Cottingham
sum rule\cite{bc} for the spin structure function $g_2(x,Q^2)$.  Thus, a 
consistent framework to evaluate the quantity $\sigma_{TS}$ is
established. In section 4,  we evaluate the longitudinal amplitude 
$S_{\frac 12}$ with the transition operator that generates the sum rule
for the quantity $\sigma_{TS}$, which has not been done systematically
in the literature.   The numerical results  for the spin structure 
functions $g_1(x,Q^2)$ and $g_2(x,Q^2)$ in the small $Q^2$ region  are 
also shown in section 4.  Our results show that the effects of the quantity 
$\sigma_{TS}$ on the spin structure functions are important in the 
resonance region.  Finally, the conclusion is given in Section 5.

\section*{2. The Sum Rule For The Quantity $\sigma_{TS}$}

Because  the spin structure functions
of the nucleon are usually measured above the pion photoproduction 
threshold, the sum rule for the quantity $\sigma_{TS}$ can be 
formulated as\cite{zz93}
\begin{equation}\label{4}
\int_0^1(g_1(x,Q^2=0)+g_2(x,Q^2=0))dx=\lim_{Q^2\to 0}
\frac {M\omega_{th}}{4\pi^2 
\alpha}\int_{\omega_{th}}^{\infty}\sigma_{TS} \frac {d\omega}{\sqrt{Q^2}}.
\end{equation}
The cross section $\sigma_{TS}$ in Eq. \ref{4} can be expressed in terms
of the transverse and longitudinal helicity amplitudes, which is
\begin{eqnarray}\label{6}
\sigma_{TS}
=\frac{\pi}{\sqrt{2}}\sum_{f>i}\{
\langle i,\frac 12\mid H_{l}^{*}\mid f\rangle \langle f\mid H_{t}\mid 
i,-\frac 12\rangle \nonumber \\ +\langle i,-\frac 12
\mid H_{t}^{*}\mid f\rangle \langle f,\mid H_{l}\mid i,\frac 12\rangle \}
\delta(\omega-\omega_f),
\end{eqnarray}
where $H_t$ is the transverse transition operator, and the longitudinal
 transition operator $H_l$ is defined as
\begin{equation}\label{7} 
H_l=\epsilon_{0}J_0-\epsilon_{3}J_3.
\end{equation}
Using the gauge invariant condition, $k_{\mu}J^{\mu}=k_{\mu}
\epsilon^{\mu}=0$, and choosing the longitudinal polarization vector 
$\epsilon_\mu$ as
\begin{equation}\label{8}
\epsilon_{\mu}^{L}=\{\epsilon_{0},0,0,\epsilon_{3}\}
=\left \{\frac{k_{3}}{\sqrt{Q^{2}}},0,0,\frac{\omega}{\sqrt{Q^{2}}}
\right \},
\end{equation}
we have
\begin{equation}\label{9}
\langle f\mid H_l\mid i\rangle =\frac{\sqrt{Q^2}}{\omega}
\langle f\mid J_3\mid i\rangle,
\end{equation}
or
\begin{equation}\label{377}
\langle f|H_l|i\rangle = \frac {\sqrt{Q^2}}{k} \langle f|J_0|i\rangle.
\end{equation}
Both Eqs. \ref{9} and \ref{377} can be used in the
 evaluation of the quantity $\sigma_{TS}$  because of the current
conservation.  It can be shown that the quantity $\sigma_{TS}$ is 
independent of which longitudinal current being used.   Consequently,
the sum rule for the quantity $\sigma_{TS}$ does not depend on the choice 
of the longitudinal current as well, as long as it is gauge invariant.

Substitute Eq. \ref{6} into Eq. \ref{4}, we have
\begin{eqnarray}\label{10}
\int_{\omega_{th}}^{\infty}\sigma_{TS}\frac{d\omega}{\sqrt{Q^2}}
=\frac{\pi}{\sqrt{2}\omega}\sum_{f>i}\big \{
\langle i,\frac 12\mid J_3^{*}\mid f\rangle \langle f\mid 
H_{t}\mid i,-\frac 12\rangle  \nonumber \\ +\langle i,-\frac 12\mid 
H_{t}^{*}\mid f\rangle \langle f\mid J_3\mid i,\frac 12\rangle\big \}.
\end{eqnarray}
The operator $H_t$ in Eq. \ref{10} is also responsible to generate
the DHG\cite{dhg} sum rule for the transverse cross section 
$\sigma_{\frac 12}-\sigma_{\frac 32}$ in the real photon limit.
Thus, the consistency requires that the same $H_t$ in Eq. \ref{11}
 should be also used in deriving Eq. \ref{35}.  
Following the same procedure as that in Ref. \cite{zpli}, we rewrite 
Eq. \ref{11} by separating the center of mass motion from the internal
motion:
\begin{equation}\label{12}
H_t=\sqrt{\frac {\omega}2} (h^c+h^p),
\end{equation}
where
\begin{equation}\label{13}
h^c=i\sum_{j}\left \{e_{j}\vec{R}\cdot\vec{\epsilon}
-\frac{1}{4M_{T}}(\frac{2e_{j}}{m_{j}}-\frac{e_{T}}{M_{T}})
\vec{\sigma}_{j}\cdot(\vec{\epsilon}\times\vec{P}_{T})\right \} 
+\hat {\mu}^c,
\end{equation}
and
\begin{equation}\label{14}
h^p = i\sum_{j}\left \{
e_{j}\vec{\epsilon}\cdot(\vec{r}_{j}-\vec{R})
-\frac{1}{4}(\frac{e_{j}}{m_{j}}-\frac{e_{T}}{M_{T}})
\vec{\sigma}_{j}\cdot(\vec{\epsilon}\times 
(\frac{\vec{p}_{j}}{m_{j}}-\frac{\vec{P}_{T}}{M_{T}}))\right \} 
+\hat{\mu}^p,
\end{equation}
where $\epsilon=-\frac{1}{\sqrt{2}}(1,i,0)$ is the transverse polarized
vector of photons. The last terms $\hat{\mu}^c$ and $\hat {\mu}^p$ in Eqs. 
\ref{13} and \ref{14} correspond to the second term in Eq. \ref{11}.  Their
 contributions to $\sigma_{TS}$ is  more complicated, and will be
evaluated separately.

The longitudinal transition operator $J_3$ can be obtained by simply 
replacing the polarization vector $\vec \epsilon$ with the vector
$\hat {\vec k}$, and the
separation of the center of mass  from the internal motions for the
longitudinal transition  is therefore easy to follow
\begin{equation}\label{15}
J_3=\sqrt{\frac {\omega}2} (j^c+j^p),
\end{equation}
where
\begin{equation}\label{16}
j^c=\sum_{j}\left \{ ie_{j}\vec{R}\cdot\hat{\vec{k}}
+\frac{i}{4M_{T}}(\frac{e_{T}}{M_{T}}
-\frac{2e_{j}}{m_{j}})\vec{\sigma}_{j}\cdot(\hat{\vec{k}}
\times\vec{P}_{T})\right \},
\end{equation}
and
\begin{equation}\label{17}
j^p=\sum_{j}\left \{
i\hat {\vec{k}}\cdot (\vec{r}_{j}-\vec{R})e_{j}
+\frac{i}{4}(\frac{e_{T}}{M_{T}}-\frac{e_{j}}{m_{j}})
\vec{\sigma}_{j}\cdot (\hat {\vec{k}}\times 
(\frac{\vec{p}_{j}}{m_{j}}-\frac{\vec{P}_{T}}{M_{T}}))\right\}.
\end{equation}

Now we are in the position to derive Eq. \ref{35} from the 
constituent quark model.  Substituting Eqs. \ref{12} and \ref{15} into 
Eq. \ref{10},  we have
\begin{eqnarray}\label{18}
\int_{\omega_{th}}^{\infty}\sigma_{TS}\frac{d\omega}{\sqrt{Q^2}}
=\frac{4\pi^2\alpha}{2\sqrt{2}}\sum_{f>i}\big \{
\langle i,1/2\mid {j^c}^{*}\mid f \rangle \langle f\mid 
h^c\mid i,-1/2 \rangle \nonumber \\
+\langle i,-1/2\mid {h^c}^{*}\mid f \rangle \langle 
f \mid j^c\mid i,1/2 \rangle  \nonumber \\ 
+\langle i,1/2\mid {j^p}^{*}\mid f \rangle \langle f\mid 
h^p\mid i,-1/2 \rangle \nonumber \\
+ \langle i,-1/2\mid {h^p}^{*}\mid f \rangle \langle 
f\mid j^p\mid i,1/2 \rangle \big \} ,
\end{eqnarray}
where the charge $e^2$ has been written as $4\pi \alpha$ explicitly
so that the total charge for protons becomes $1$ instead of $e$. 
Using the closure relation, Eq. \ref{18} becomes
\begin{eqnarray}\label{19}
\sum_{f>i}\big \{\langle i,\frac 12 \mid {j^c}^{*}\mid f 
\rangle \langle f\mid h^c\mid i,-\frac 12 \rangle 
+\langle i,-\frac 12\mid {h^c}^*\mid f
\rangle \langle f\mid j^c\mid i,\frac 12 \rangle  \nonumber \\
+\langle i,\frac 12\mid {j^p}^{*}\mid f \rangle \langle f\mid 
h^p\mid i,-\frac 12 \rangle
+ \langle i,-\frac 12\mid {h^p}^{*}\mid f
\rangle \langle f\mid j^p\mid i,\frac 12 \rangle \big \}  \nonumber \\ 
= \langle i, \frac 12 \mid {j^c}^*h^c+ {j^p}^*h^p\mid i,-\frac 12 \rangle  
+ \langle i, -\frac 12 \mid {h^c}^* j^c +{h^p}^{*} j^p \mid i, 
\frac 12\rangle  \nonumber \\ 
-\langle i,\frac 12\mid {j^c}^{*}\mid i\rangle \langle i\mid 
h^c\mid i,-\frac 12 \rangle 
-\langle i,-\frac 12\mid {h^c}^*\mid i
\rangle \langle i,\mid j^c\mid i,\frac 12 \rangle  .
\end{eqnarray} 
We first consider the correlations between the longitudinal transition 
operators, $j^c$($j^p$), and the first term $h^c-\mu^c$($h^p-\mu^p$)
 in Eq. \ref{13} (\ref{14}). In the Appendix  we show that
\begin{eqnarray}\label{20}
 \langle i, \frac 12 \mid {j^c}^*(h^c-\hat {\mu}^c)\mid i -\frac 12\rangle +
\langle i, -\frac 12 \mid ({h^c}^*-{\hat {\mu^c}}^{*}) j^c \mid i, \frac 12
\rangle \nonumber \\
=-\langle i, \frac 12\mid \sum_{j}
\sigma^{+}_{j}\frac{e_{T}}{2M_{T}}
\left (\frac{2e_{j}}{m_{j}}-\frac{e_{T}}{M_{T}}\right )\mid i,-\frac 12
 \rangle\sqrt{2} ,
\end{eqnarray}
and similarly
\begin{eqnarray}\label{21}
 \langle i, \frac 12 \mid {j^p}^*(h^p-\hat {\mu}^p)\mid i, 
-\frac 12\rangle + \langle i, -1/2 \mid ({h^p}^*-{\hat {\mu^p}}^{*}) 
j^p \mid i, \frac 12 \rangle   \nonumber \\ 
=\langle i, \frac 12 \mid \sum_{j}
\sigma^{+}_{j}\left (\frac{e_{j}}{m_{j}}-\frac{e_{T}}{M_{T}}\right )
\left (\frac{e_{T}}{2M_{T}}-\frac{e_{j}}{2m_{j}}\right )
\mid i,-\frac 12 \rangle \sqrt{2}.
\end{eqnarray}

We turn to the correlation between the magnetic term $\hat {\mu}$ 
and the longitudinal transition operator $j$.  The leading term for
the operator $\mu$ is 
\begin{equation}\label{221}
\mu^c_0=\sum_j\frac{e_{j}}{2m_{j}}
\vec{\sigma}_{j}\cdot (\vec{\epsilon}\times
\hat{\vec k}),
\end{equation}
and
\begin{equation}\label{222}
\mu^p_0=0.
\end{equation}
The correlation between $\mu^c_0$ and $j^c$ gives
\begin{eqnarray}\label{22}
\langle i,\frac 12\mid {j}^{c*}\hat {\mu}^{c}_{0}\mid i,-\frac 12 \rangle 
+\langle i,-\frac 12 \mid {\hat {\mu}}^{c*}_{0}j^{c}\mid i,\frac 12 \rangle 
\nonumber \\
=\sum_j\bigg \{\langle i,\frac 12 \mid  e_T\vec R \cdot \hat
 {\vec{k}}\frac{e_{j}}{2m_{j}}\vec{\sigma}_{j}\cdot 
(\vec{\epsilon}\times \hat{\vec k}) \mid i, -\frac 12\rangle
 \nonumber \\ +
\langle i, -\frac 12 \mid \frac{e_{j}}{2m_{j}}
\vec{\sigma}_{j}\cdot (\vec{\epsilon}^*\times
\hat{\vec k} ) e_T\vec R \cdot \hat {\vec{k}}
\mid i \frac 12\rangle \bigg \} ,
\end{eqnarray}
Substitute the polarization $\epsilon=-\frac 1{\sqrt {2}}(1,i,0)$ and
$\epsilon^*=-\frac 1{\sqrt {2}}(1,-i,0)$ into Eq. \ref{22}, we find that
the two terms in Eq. \ref{22} cancel each other so that it
vanishes.  Thus, the nonzero contribution from the 
correlation between $\hat {\mu}$ and the longitudinal transition $j$
should come from the next order.  By expanding the photon wave function
$e^{i\vec k\cdot \vec {r}_j}$, we have 
\begin{equation}\label{23}
\hat {\mu}=\sum_j \frac{e_{j}}{2m_{j}}
\vec{\sigma}_{j}\cdot (\vec{\epsilon}\times
\vec{\hat k}) (1+i \vec k \cdot \vec {r}_j +O(k)) .
\end{equation}
Because the leading term in Eq. \ref{23} gives a zero contribution to 
the quantity $\sigma_{TS}$, we examine the second term
in Eq. \ref{23}.  In the real photon limit, we rewrite the second term
in Eq. \ref{23} as
\begin{eqnarray}\label{24}
\langle f \mid 
i \sum_j \frac{e_{j}}{2m_{j}}
\vec{\sigma}_{j}\cdot (\vec{\epsilon}\times
\vec{\hat k})  \vec k \cdot \vec {r}_j \mid i \rangle 
=i \langle f \mid [H,
\sum_j \frac{e_{j}}{2m_{j}}
\vec{\sigma}_{j}\cdot (\vec{\epsilon}\times
\vec{\hat k})  \hat {\vec k} \cdot \vec {r}_j] \mid i \rangle 
\nonumber \\
=\langle f \mid \sum_j \frac{e_{j}}{2m_{j}}
\vec{\sigma}_{j}\cdot (\vec{\epsilon}\times
\vec{\hat k})  \hat {\vec k}
 \cdot \frac {\vec {p}_j}{m_j} \mid i \rangle,
\end{eqnarray}
so that the closure relation could be used, because the transition
operator has no explicit dependence on the transition energy $\omega$. 
The operator $H$ in Eq. \ref{24} is the Hamiltonian of the system; 
\begin{equation}\label{25}
H=\sum_j \frac { {\vec p}_j^2}{2m_j} + \sum_{i,j} V_{ij}({\vec r}_i-
{\vec r}_j).
\end{equation}
Therefore, $\hat {\mu}^c$ and $\hat {\mu}^p$ in Eqs. \ref{13} and \ref{14}
are 
\begin{equation}\label{26}
\hat {\mu}^c_1 =\sum_j
\frac{e_{j}}{2m_{j}}
\vec{\sigma}_{j}\cdot (\vec{\epsilon}\times
\vec{\hat k})\vec{\hat k}\cdot \frac{\vec{P}_{T}}{M_{T}}
\end{equation}
and
\begin{equation}\label{27}
\hat {\mu}^p_1=\sum_j \frac{e_j}{2m_j}
\vec{\sigma}_{j}\cdot (\vec{\epsilon}\times
\vec{\hat k})\vec{\hat k}\cdot (\frac{\vec{p}_{j}}{m_{j}}
-\frac{\vec{P}_{T}}{M_{T}}).
\end{equation}
The correlation between $\hat {\mu}^{c,p}_1$ and $j^{c,p}$ gives
\begin{eqnarray}\label{29}
\langle i,\frac 12 \mid  {j^c}^* \hat {\mu}^c_1 + 
{j^p}^* \hat {\mu}^p_1 \mid i,-\frac 12\rangle +
\langle i,-\frac 12 \mid  {\hat {\mu^c}}^{*}_1 j^c + {\hat {\mu^p}}^{*}_1
j^p\mid i,\frac 12\rangle  \\ \nonumber 
=\langle i, \frac 12\mid \sum_{j}
\sigma^{+}_{j}\left [\frac {e_T}{2M_T}
\frac{e_j}{m_j}+\left (\frac{e_j}{m_j}-\frac {e_T}{M_T}\right )
\frac{e_j}{2m_j} \right ]\mid i,-\frac 12 \rangle\sqrt{2} .
\end{eqnarray}
This shows that the nonzero contributions to the correlation between 
the magnetic transition $\hat {\mu}$ and the longitudinal transition 
operator $j$ come from the higher order expansion in $\hat {\mu}$, 
this  feature  does not exist in the transverse
transverse correlations that leads to the DHG sum rule.

Therefore, combine Eqs. \ref{21}, \ref{22} and \ref{29}, we have 
\begin{equation}\label{30}
\langle i,\frac 12 \mid  {j^c}^* h^c + 
{j^p}^* h^p \mid i, -\frac 12\rangle +
\langle i, -\frac 12 \mid  {h^c}^{*} j^c + {h^p}^{*}j^p
\mid i, \frac 12\rangle =0.
\end{equation}
That is, the sum rule for the quantity $\sigma_{TS}$
is only determined by the static properties of ground states:
\begin{eqnarray}\label{31}
\int_{\omega_{th}}^{\infty}\sigma_{TS}\frac{d\omega}{\sqrt{Q^2}}
=-\frac {4\pi^2\alpha}{2\sqrt{2}}\bigg \{ 
\langle i,\frac 12\mid {j^c}^{*}\mid i\rangle \langle i\mid 
h^c\mid i,-\frac 12 \rangle \\ \nonumber
+\langle i,-\frac 12\mid {h^c}^{*}\mid i\rangle \langle 
i\mid j^c\mid i,\frac 12\rangle\bigg \} .
\end{eqnarray}
This is a general feature for the sum rules of both $g_1(x,Q^2)$ and
$g_2(x,Q^2)$; the sum rules in real photon limit do not depend on the 
internal structure of the nucleon so that it behaves like an elementary
particle in the low energy limit.
Using the relation
\begin{equation}\label{32}
\langle i | \sum_j \frac {e_j}{2m_j} {\vec \sigma}_j | i\rangle 
=\mu \langle i | {\vec \sigma}_T| i\rangle
\end{equation}
where ${\vec \sigma}_T$ is the total spin operator of a many-body system,
we have 
\begin{equation}\label{33}
\langle i | h^c| i\rangle = \langle i | H^c| i\rangle
\end{equation}
and
\begin{equation}\label{34}
\langle i | j^c| i\rangle = \langle i |J^c|i\rangle,
\end{equation}
where
\begin{equation}\label{333}
H^c=i\bigg \{e_T {\vec R}\cdot {\vec \epsilon}+\mu
{\vec \sigma}_T \cdot ({\vec  \epsilon} \times  \hat {\vec k})
\frac {{\vec P}_T\cdot \hat {\vec k}}{M_T}
-\frac 1{2M_T}\left (2\mu -\frac {e_T}{2M_T}\right )
{\vec \sigma}_T\cdot ({\vec \epsilon} \times {\vec P}_T)\bigg \}
\end{equation}
and 
\begin{equation}\label{344}
J^c=i\bigg \{e_T {\vec R}\cdot {\vec \epsilon}
-\frac 1{2M_T}\left (2\mu -\frac {e_T}{2M_T}\right )
{\vec \sigma}_T\cdot ({\vec \epsilon} \times {\vec P}_T)\bigg \}.
\end{equation}
Thus, the closure relation can be used because the operators 
$H^c$ and $J^c$ do not connect the ground state with the excited
states:
\begin{eqnarray}\label{355}
\langle i, \frac 12 |{j^c}^*|i\rangle \langle i|h^c| i,-\frac 12\rangle
+\langle i,-\frac 12|{h^c}^*|i\rangle \langle i|j^c| i,\frac 12 \rangle
\nonumber \\
=\langle i, \frac 12 | {J^c}^*H^c| i,-\frac 12\rangle + \langle i, 
-\frac 12 | {H^c}^*J^c|i,\frac 12\rangle  
\nonumber \\
=\frac {\sqrt{2}}{2M^2_T} e_T\kappa ,
\end{eqnarray}
 which leads to the sum rule in Eq. \ref{35}.
Consequently, the sum rule for the spin structure function $g_{2}$ 
in the real photon limit is just a linear combination of Eq. \ref{35}
and the DHG sum rule:
\begin{equation}\label{37}
\lim_{Q^2\to 0}
\int_{0}^{1}dxg_{2}(x,Q^{2})=\frac {\omega_{th}}{M_T}\frac{\kappa 
(\kappa +e_{T})}{4}.
\end{equation}

This shows that the sum rules for both $g_{1}(x,Q^2)$ and $g_2(x,Q^2)$
in the real photon limit can be derived consistently from the same set
of the transition operators in the quark model.  It also highlights
the importance of the spin-orbit interaction and the nonadditive term
in both transverse and longitudinal transition operators $H_t$ and $J_3$.
In the next section, we will give an intuitive proof that the same is
also true for the sum rules in the large $Q^2$ limit.

\section*{3. The Extension To The Large $Q^2$ limit}

In the case of $Q^2\not= 0$, an extra term is generated from the
transverse operator $h=h^c+h^p$ so that
\begin{equation}\label{38}
h=h_0+h_1,
\end{equation}
where $h_0$ represents the transition operator $h$ at $Q^2=0$, and
\begin{equation}\label{39}
h_1=\sum_j\frac {Q^2}{\omega^2}\frac{e_{j}}{2m_{j}}
\vec{\sigma}_{j}\cdot (\vec{\epsilon}\times\vec{\hat k})  \hat {\vec k}
 \cdot \frac {\vec {p}_j}{m_j},
\end{equation}
while the longitudinal operator $j=j^c+j^p$ remains the same. 
Eq. \ref{30} shows that the correlation between $h_0$
and $j$ is zero for the inclusive processes, thus only the correlation 
between $h_1$ and $j$ needs to be investigated. 
Note that the Bjorken scaling variable $x_j$ 
is related to the photon energy and the mass of partons\cite{zpli}:
\begin{equation}\label{40}
x_{j}=\frac{Q^{2}}{2M_{T}\omega}=\frac{m_{j}}{M_{T}}
\end{equation}
in the large $Q^2$ limit. The operator $h_1$ can be written as
\begin{equation}\label{41}
h_1=\sum_j\frac {2}{Q^2}e_{j}
\vec{\sigma}_{j}\cdot (\vec{\epsilon}\times\hat{\vec k})  \hat {\vec k}
 \cdot {\vec {p}_j}.
\end{equation}
The correlation between $h_1$ and $j$ gives
\begin{eqnarray}\label{42}
\langle i,\frac 12 \mid  {j}^* h_1\mid i,-\frac 12\rangle +
\langle i,-\frac 12 \mid  h_1^*j \mid i,\frac 12\rangle \nonumber \\
=\frac {2\sqrt{2}}{Q^2}\langle i, \frac 12|\sum_je_j^2\sigma_j^+|i, -\frac 12
\rangle .
\end{eqnarray}
Therefore, we have 
\begin{equation}\label{43}
\lim_{Q^2\to \infty}
\int_{\omega_{th}}^{\infty}\sigma_{TS}\frac{d\omega}{\sqrt{Q^2}}
=\frac {4\pi^2\alpha}{Q^2} \langle i, \frac 12|\sum_je_j^2\sigma_j^+|i,
 -\frac 12\rangle .
\end{equation}
A similar procedure in the large $Q^2$ extension of the DHG sum rule 
gives
\begin{equation}\label{44}
\int_{\omega_{th}}^{\infty}(\sigma_{\frac 12}-\sigma_{\frac 32})
\frac{d\omega}{\omega}
=\frac {4\pi^2\alpha}{Q^2} \langle i|\sum_j e_j^2\sigma_j^z|i\rangle_{P-A}.
\end{equation}
 Combining Eqs. \ref{43}
and \ref{44} gives the well known Burkhardt-Cottingham(BC) sum 
rule\cite{bc} for the spin structure function $g_2$,
\begin{equation}\label{45}
\int_0^1 g_2(x)dx=0.
\end{equation}
Therefore, the sum rules for both $g_1(x,Q^2)$ and $g_2(x,Q^2)$
can be obtained from the same set of electromagnetic
interactions in Eqs. \ref{11} and \ref{15} for a many body system.
It shows that the transition of
the spin dependent sum rules (both DHG sum rule and Eq. \ref{35}) 
from the real photon limit to the large $Q^2$ limit is an evolution
from an exclusive, coherent elastic scattering to an inclusive, 
incoherent deep-inelastic scattering of a many-body system.
Moreover, by reproducing the spin-dependent sum rules in the real
photon and large $Q^2$ limits, we are able to establish a framework
to evaluate the spin structure functions of nucleons in the finite
$Q^2$ region, where the contributions from baryon resonances are 
important, and the quark model has been very successful.

\section*{4. The evaluation of the
spin dependent sum rules in the low $Q^2$ region}

The numerical studies of the quantity $\sigma_{TS}$ require the 
evaluations of the transverse helicity amplitude $A_{\frac 12}$ and the 
longitudinal amplitude $S_{\frac 12}$.  The helicity amplitude 
$A_{\frac 12}$ has been calculated\cite{cl91} by using
 the transition operator in Eq. \ref{11} that generates the DHG sum rule.  
Thus, only the longitudinal amplitude $S_{\frac 12}$ needs to be 
evaluated.  Following Eq. \ref{377}, the longitudinal transition 
amplitude $S_{\frac 12}$ is
\begin{equation}\label{46}
S_{\frac 12}=\langle f|J_0|i\rangle, 
\end{equation}
and we have the longitudinal transition operator\cite{cl93}
\begin{eqnarray}\label{47}
J_{0}&=&\sqrt{\frac{1}{2\omega}}\{
\sum_{j}(e_{j}+\frac{ie_{j}}{4m_{j}^{2}}\vec{k}\cdot (\vec{\sigma}_{j}
\times\vec{p}_{j}))e^{i\vec{k}\cdot\vec{r}_{j}}) \nonumber  \\
&&-\sum_{j<l}\frac{i}{4M_{T}}
(\frac{\vec{\sigma}_{j}}{m_{j}}-\frac{\vec{\sigma}_{l}}{m_{l}})
\cdot
(e_{j}\vec{k}\times\vec{p}_{l}e^{i\vec{k}\cdot\vec{r}_{j}}
-e_{l}\vec{k}\times\vec{p}_{j}e^{i\vec{k}\cdot\vec{r}_{l}})\},
\end{eqnarray}
where the second and third terms are the spin-orbit and nonadditive terms
that represents the relativistic corrections to the leading charge operator.
The study in previous sections clearly shows that the spin-orbit and the 
nonadditive terms are crucial in reproducing the sum rule for the quantity 
$\sigma_{TS}$.  

Because the longitudinal amplitude $S_{\frac 12}$ of baryon resonances has 
not been systematically calculated with the transition operator $J_0$ 
in Eq. \ref{47},  we show the analytical expressions of the longitudinal 
amplitudes $S_{\frac 12}$ between the nucleon and baryon resonances in 
$SU(6)\otimes O(3)$ symmetry limit in Table 1. The evaluation of the $Q^2$
dependence of the longitudinal amplitudes $S_{\frac 12}$ follows the
procedure of Foster and Hughes\cite{foster}, and the
the longitudinal amplitudes $S_{\frac 12}$ as a function of $Q^2$
 for the resonance $S_{11}(1535)$, $D_{13}(1520)$ and $F_{15}(1688)$ are 
shown in Fig. 1.  These results are in better agreement with the analysis
by Gerhardt\cite{gerh} than the previous calculations\cite{warns}, 
who extracted the longitudinal amplitudes from the electroproduction data.
The numerical results in Fig. 1 shows that the longitudinal amplitudes
are quite large in the low $Q^2$ region, thus suggest that they play 
a significantly role in the spin structure functions of nucleon in
the low $Q^2$ region.

The resonance contributions to the sum rules of the spin structure 
functions can be expressed in terms of the helicity amplitudes, 
$A_{\frac 12}$ and $A_{\frac 32}$, and the longitudinal amplitudes 
$S_{\frac 12}$:
\begin{equation}\label{48}
\int g_{1}(x,Q^{2})dx=\sum_{R} K.E.
\left [ |A_{\frac 12}^R|^2-|A_{\frac 32}^R|^2
+\frac{Q^2}{\sqrt{2}\omega k}({S_{\frac 12}^R}^*A_{\frac 12}^R
+{A_{\frac 12}^R}^*S_{\frac 12}^R)\right ]
\end{equation}
and
\begin{equation}\label{49}
\int g_{2}(x,Q^{2})dx=\sum_R K.E.
\left [\frac{\omega}{\sqrt{2}k}({S_{\frac 12}^R}^*A_{\frac 12}^R
+{A_{\frac 12}^R}^*S_{\frac 12}^R)-\left (|A_{\frac 12}^R|^2
-|A_{\frac 32}^R|^2\right )\right ],
\end{equation}
where the kinetic factor K.E.  is
\begin{equation}\label{50}
K.E. =\frac{M\omega_{th}}{4\pi\alpha(1+\frac {Q^{2}}{\omega^{2}})\omega}
\end{equation}
and $\omega_{th}$ is given in Eq. \ref{5}.  The total width of each 
resonance is treated as zero so that the integration over the photon
energy can be approximated by a summation over all the resonances.
The background contributions from the nucleon born terms in the single
pion photoproductions are not included in Eqs. \ref{48} and Eq. \ref{49},
and they can be easily included later in a more detail studies.
Because these amplitudes are evaluated by using 
the transition operators that generate the the spin dependent sum rules,
the calculations of the spin structure function in the resonance region
become straightforward.  In Fig. 2, we show the resonance contributions
to the sum rule for $g_1(x,Q^2)$,
in which every resonance below 2 GeV is included.  The resonances 
$P_{11}(1440)$ and $P_{33}(1600)$ are treated as hybrid states\cite{zl},
and the study shows\cite{zpli93} that the $Q^2$ dependence of the transition
amplitudes for hybrid $P_{11}(1440)$ and $P_{33}(1600)$ gives a better
agreement with the existing data. The numerical result
for the $Q^2$ dependence of the integral for  the  transverse cross 
section $\sigma_{\frac 12}-\sigma_{\frac 32}$ is also shown in Fig. 2,
and it is in good agreement with a more sophisticated evaluation in Ref. 
\cite{zz93}.  The resonance contribution to the integral 
$\int g_1(x,Q^2=0)dx$ at the real photon limit is -0.121,
 which is in good agreement with the theoretical prediction
 $-\frac {\omega_{th}}{4M_T}\kappa^2$ with $k_p=1.79$ for the proton
target\cite{zz93}. This result is consistent with the conclusions of 
our previous investigation\cite{zpli}; the contributions from resonances, 
in particular the resonance $P_{33}(1232)$, dominate the DHG sum rule.
The difference between the $g_1(x,Q^2)$ sum rule and 
the contribution from the quantity $\sigma_{\frac 12}-\sigma_{\frac 32}$
shows the importance of the quantity $\sigma_{TS}$.  
It is particularly significant in the small $Q^2$ region, and the addition 
of the quantity $\sigma_{TS}$ has pushed the crossing point that 
the sum rule is zero from $0.7$ GeV$^2$ to around 0.5 GeV$^2$.     

The sum rule for the spin structure function $g_2(x,Q^2)$ in the resonance
region is shown in Fig. 3.  The resonance contributions to the sum rule 
$\int g_2(x,Q^2)$ at the real photon limit is 0.182, while Eq.\ref{37}
gives 0.192 for the proton target. This shows that the resonance 
contributions dominate the sum rule for $g_2(x,Q^2)$ in the real photon
limit as well.  The $g_2(x,Q^2)$ is only significant 
in the $Q^2\le 1$ GeV$^2$ region, and decreases very quickly 
as $Q^2$ increases.  There is also a sign change for the 
sum rule of $g_2(x,Q^2)$ at $Q^2\approx 1$ GeV$^2$.  A recent
calculation\cite{burk} in the single pion channel of pion 
photoproduction has shown a similar
behaviour,  in which only the nucleon born term is considered.
This behaviour is not consistent with the $Q^2$ dependence of the sum rule
of $g_2(x,Q^2)$ derived in Ref. \cite{schwinger,sofer}.  It may 
represent the theoretical uncertainty of the quark model calculations.   
On the other hand, it would be very interesting to see if there is a sign 
change in the experimental data.

To highlight the importance of the quantity $\sigma_{TS}$ in the 
resonance region, we present an estimate of the total sum rule for 
$g_1(x,Q^2)$ by including the contributions from outside the resonance 
region.  Following the procedure in Ref. \cite{zz93},
the total spin dependent sum rule should be written as
\begin{equation}\label{51}
\int_0^1g_1(x,Q^2)=\int_{x_r}^1+\int_0^{x_r}g_{1}(x,Q^2)dx,
\end{equation}
where 
\begin{equation}\label{52}
x_r=\frac {Q^2+2m_{\pi}M_T+m_{\pi}^2}{W_r^2+Q^2-M_T^2}
\end{equation}
with $W_r=2.0$ GeV.  The first term in Eq. \ref{51} represents
the contributions from the resonance region, it shows that the
contributions from the resonance region does not cover the whole
kinetic region from $x=0$ to $x=1$.
The second term in Eq. \ref{51} comes from the outside resonance 
region, and we showed in Ref. \cite{zz93} that this term becomes 
increasingly important as $Q^2$ increases.  Because 
there is no experimental information on the quantity $\sigma_{TS}$
outside the resonance region, one could only make a qualitative 
estimate on the second term in Eq. \ref{51}.  We show the
estimate of the $Q^2$ dependence of the spin dependent sum rule
$\int_0^1 g_{1}(x,Q^2) dx$ in Fig. 4. The contribution
from the second term is obtained from the estimate of the nonresonant
 contribution in Ref. \cite{zz93}, in which the quantity $\sigma_{TS}$
is not included.  Thus, this estimate could only be regarded as a
lower limit of the spin dependent sum rule for $g_{1}(x,Q^2)$.
Nevertheless, the effects of the quantity $\sigma_{TS}$ on the $Q^2$
dependence of the sum rule $\int_0^1 g_{1}(x,Q^2) dx$ are very important,
and it could not be neglected if the high twist term that generates the
leading $1/Q^2$ corrections to the spin structure function in the deep 
inelastic scattering region is extracted from $Q^2 \approx 1.5 \sim 2.5$
GeV$^2$ region.

\section*{\bf 5. Conclusion}

We have presented a consistent framework to investigate the spin 
structure functions of nucleon in the resonance region, which the
model independent sum rules in the real photo limit and the large
$Q^2$ limit are satisfied.  We show that the same set of 
transition operator generates both DHG sum rule for the transverse
cross section, $\sigma_{\frac 12}-\sigma_{\frac 32}$, and the sum rule 
for the quantity $\sigma_{TS}$.  
The sum rule for the quantity $\sigma_{TS}$ also provides a crucial
constraint on the longitudinal transition operator; it requires the 
longitudinal transition operator to be gauge invariant and to be expanded
to order $O(\frac {v^2}{c^2})$ consistently.  The operator in 
Eq. \ref{47} satisfies these requirements.  This clarifies some of 
the problems in the literature on the longitudinal transitions, 
although the problem of the model space truncation discussed in Ref.
\cite{muko} is not considered here.

A more quantitative calculation of the spin dependent sum rules for
both spin structure functions $g_1(x,Q^2)$ and $g_2(x,Q^2)$ are 
presented for the first time in the quark model.  Our numerical results
 indicate that the
effects of the quantity $\sigma_{TS}$ are very important in small 
$Q^2$ region,  which certainly can be tested in the future experiments
at CEBAF\cite{cpr}.

\subsection*{Acknowledgements}
Discussions with V.Burkert are gratefully acknowledged.  This work
is support in part by U.S. National Science Foundation PHY-9023586.

\section*{Appendix}
The terms that contribute to the spin flip in Eq. \ref{20} are
\begin{eqnarray}\label{60}
\langle i, \frac 12 |{j^c}^*(h^c-\hat {\mu^c})|i,-\frac 12\rangle +
\langle i, -\frac 12 |(h^c-\hat {\mu^c})^*j^c|i,\frac 12\rangle =
\nonumber \\
\sum_j \langle i,\frac 12| \frac {e_T}{4M_T}\left (\frac {e_T}{M_T}-
\frac {2e_j}{m_j}\right ) \vec {\sigma}_j \cdot (\hat{\vec k}\times
{\vec P}_T) \vec R\cdot \vec \epsilon |i, -\frac 12\rangle   \nonumber \\
+\sum_j \langle i,-\frac 12| \frac {e_T}{4M_T}\left (\frac {e_T}{M_T}-
\frac {2e_j}{m_j}\right )\vec R\cdot \vec \epsilon^*
 \vec {\sigma}_j \cdot (\hat{\vec k}\times
{\vec P}_T) |i, \frac 12\rangle \nonumber \\ 
+\sum_j \langle i,\frac 12| \frac {e_T}{4M_T}\left (\frac {e_T}{M_T}-
\frac {2e_j}{m_j}\right )\vec R\cdot\hat{\vec k} 
\vec {\sigma}_j \cdot (\vec \epsilon\times
{\vec P}_T)|i, -\frac 12\rangle  \nonumber \\
+\sum_j \langle i,-\frac 12| \frac {e_T}{4M_T}\left (\frac {e_T}{M_T}-
\frac {2e_j}{m_j}\right ) \vec {\sigma}_j \cdot (\vec \epsilon^*\times
{\vec P}_T)\vec R\cdot\hat{\vec k}|i, \frac 12\rangle
\end{eqnarray}
Lets consider the terms proportional to $\vec R\cdot \vec \epsilon$ in
Eq. \ref{60}. The product $ \vec {\sigma}_j \cdot (\hat{\vec k}\times
{\vec P}_T)$ can be written as
\begin{equation}\label{61}
 \vec {\sigma}_j \cdot (\hat{\vec k}\times
{\vec P}_T) =-i \sigma^+_j (P_x-iP_y)+i\sigma^-_j(P_x+iP_y),
\end{equation} 
where $\sigma^{\pm}=\frac 12 (\sigma_x\pm i\sigma_y)$.
Substitute $\epsilon=-\frac 1{\sqrt{2}}(1,i,0)$ into Eq. \ref{60}, we
have
\begin{eqnarray}\label{62}
\sum_j \langle i,\frac 12| \frac {e_T}{4M_T}\left (\frac {e_T}{M_T}-
\frac {2e_j}{m_j}\right )\vec {\sigma}_j \cdot (\hat{\vec k}\times
{\vec P}_T) \vec R\cdot \vec \epsilon|i, -\frac 12\rangle \nonumber \\
+\sum_j \langle i,-\frac 12| \frac {e_T}{4M_T}\left (\frac {e_T}{M_T}-
\frac {2e_j}{m_j}\right )\vec R\cdot \vec \epsilon^*
 \vec {\sigma}_j \cdot (\hat{\vec k}\times
{\vec P}_T) |i, \frac 12\rangle \nonumber \\
=\sum_j \langle i,\frac 12| \frac {e_T}{4M_T}\left (\frac {e_T}{M_T}-
\frac {2e_j}{m_j}\right ){\sigma}^+_j\frac i{\sqrt{2}}P^-
R^+|i, -\frac 12\rangle \nonumber \\
-\sum_j \langle i,-\frac 12| \frac {e_T}{4M_T}\left (\frac {e_T}{M_T}-
\frac {2e_j}{m_j}\right )\sigma_j^-\frac i{\sqrt{2}}
R^-P^+|i, \frac 12\rangle,
\end{eqnarray}
where $P^{\pm}=P_x\pm iP_y$ and $R^{\pm}=R_x\pm iR_y$.
Notice that for the total 1/2 initial and final states
\begin{equation}\label{63}
\langle i, \frac 12|\sigma_j^+| i,-\frac 12\rangle = 
\langle i, -\frac 12|\sigma_j^-| i,\frac 12\rangle 
\end{equation}
in our convention for the Pauli matrix $\sigma^{\pm}$ and the
spin wave functions.  Eq. \ref{62} becomes
\begin{eqnarray}\label{64} 
\sum_j \langle i,\frac 12| \frac {e_T}{4M_T}\left (\frac {e_T}{M_T}-
\frac {2e_j}{m_j}\right ){\sigma}^+_j\frac i{\sqrt{2}}\left [P^-R^+
-R^-P^+\right ]|i, -\frac 12\rangle \nonumber \\
=\sum_j \langle i,\frac 12| \frac {e_T}{4M_T}\left (\frac {e_T}{M_T}-
\frac {2e_j}{m_j}\right ){\sigma}^+_j\sqrt{2}\left [1+iR_yP_x-iR_xP_y
\right ]|i, -\frac 12\rangle
\end{eqnarray}
where the term $R_yP_x-R_xP_y$ is an angular momentum operator for the
center of mass motions of nucleons, which is zero in this process.  
Thus, we have
\begin{eqnarray}\label{65}
\sum_j \langle i,\frac 12| \frac {e_T}{4M_T}\left (\frac {e_T}{M_T}-
\frac {2e_j}{m_j}\right )\vec {\sigma}_j \cdot (\hat{\vec k}\times
{\vec P}_T) \vec R\cdot \vec \epsilon|i, -\frac 12\rangle \nonumber \\
+\sum_j \langle i,-\frac 12| \frac {e_T}{4M_T}\left (\frac {e_T}{M_T}-
\frac {2e_j}{m_j}\right )\vec R\cdot \vec \epsilon^*
 \vec {\sigma}_j \cdot (\hat{\vec k}\times
{\vec P}_T) |i, \frac 12\rangle \nonumber \\
=\sum_j \langle i,\frac 12| \frac {e_T}{4M_T}\left (\frac {e_T}{M_T}-
\frac {2e_j}{m_j}\right ){\sigma}^+_j\sqrt{2}|i, -\frac 12\rangle.
\end{eqnarray}

Taking the same procedure, we have
\begin{eqnarray}\label{66}
\sum_j \langle i,\frac 12| \frac {e_T}{4M_T}\left (\frac {e_T}{M_T}-
\frac {2e_j}{m_j}\right )\vec R\cdot\hat{\vec k} 
\vec {\sigma}_j \cdot (\vec \epsilon\times
{\vec P}_T)|i, -\frac 12\rangle \nonumber \\
+\sum_j \langle i,-\frac 12| \frac {e_T}{4M_T}\left (\frac {e_T}{M_T}-
\frac {2e_j}{m_j}\right ) \vec {\sigma}_j \cdot (\vec \epsilon^*\times
{\vec P}_T)\vec R\cdot\hat{\vec k}|i, \frac 12\rangle \nonumber \\
=\sum_j \langle i,\frac 12| \frac {e_T}{4M_T}\left (\frac {e_T}{M_T}-
\frac {2e_j}{m_j}\right ){\sigma}^+_ji{\sqrt{2}}(P_zR_z-R_zP_z)
|i, -\frac 12\rangle \nonumber \\
=\sum_j \langle i,\frac 12| \frac {e_T}{4M_T}\left (\frac {e_T}{M_T}-
\frac {2e_j}{m_j}\right ){\sigma}^+_j\sqrt{2}|i, -\frac 12\rangle.
\end{eqnarray}
Combining Eq. \ref{65} and Eq. \ref{66} gives Eq. \ref{20}.

\newpage

\begin{table} {Table 1: Transition matrix elements between the nucleon and
baryon resonances in the $SU(6)\otimes O(3)$ symmetry limit.  The full 
matrix elements are obtained by multiplying the entries in this table by a
factor $\sqrt{\frac {2\pi}{k_0}} 2\mu m_q e^{-\frac {{\bf k}^2}{6\alpha^2}}$,
and $S_{\frac 12}^n=S_{\frac 12}^p$ for $\Delta$ states.}
\\ [1ex]
\begin{tabular}{rrcll}\hline \hline
Multiplet & States &  & Proton & Neutron  \\ \hline  
$[70,1^{-}]_1$ & $N(^2P_M){\frac 12}^{-1}$ & & $\frac 1{3\sqrt {2}}\frac 
{|\bf k|}{\alpha}\left (1+\frac {\alpha^2}{6m_q^2}\right )$ & $-\frac 1
{3\sqrt {2}}\frac {|\bf k|}{\alpha} \left (1+\frac {\alpha^2}{6m_q^2}\right )$
\\ [1ex]
& $N(^2P_M){\frac 32}^{-1}$ & & $-\frac 13 \frac {|\bf k|}{\alpha}\left
 (1-\frac {\alpha^2}{12m^2_q}\right )$ & $\frac 13\frac {|\bf k|}{\alpha}
\left (1-\frac {\alpha^2}{12m^2_q}\right )$ \\ [1ex]
& $N(^4P_M){\frac 12}^{-1}$ & & $\frac 1{36\sqrt{2}} \frac {\alpha |{\bf k}|}
{m^2_q}$ & $-\frac 1{108\sqrt{2}} \frac {\alpha |{\bf k}|}
{m^2_q}$ \\ [1ex]
& $N(^4P_M){\frac 32}^{-1}$ & & $\frac 1{9\sqrt{10}} \frac {\alpha |{\bf k}|}
{m^2_q}$ & $-\frac {5}{27\sqrt{10}} \frac {\alpha |{\bf k}|}
{m^2_q}$ \\ [1ex]
& $N(^4P_M){\frac 52}^{-1}$ & & $\frac 1{12\sqrt{10}} \frac {\alpha |{\bf k}|}
{m^2_q}$ & $-\frac {5}{36\sqrt{10}} \frac {\alpha |{\bf k}|}
{m^2_q}$ \\ [1ex]
& $\Delta(^2P_M){\frac 12}^{-1}$ & & $-\frac 1{3\sqrt{2}} \frac {|{\bf k}|}
{\alpha}\left (1-\frac {\alpha^2}{6m^2_q}\right )$ & 
 \\ [1ex]
& $\Delta(^2P_M){\frac 32}^{-1}$ & & $\frac 1{3} \frac {|{\bf k}|}
{\alpha}\left (1+\frac {\alpha^2}{12m^2_q}\right )$ & 
\\ [1ex]
$[56,0^+]_2$ & $N(^2S_{S^\prime}){\frac 12}^+$ & &
$-\frac 1{3\sqrt{6}}\frac {{\bf k}^2}{\alpha^2}$ & 0 \\ [1ex]
& $\Delta(^4S_{S^\prime}){\frac 32}^+$ & & 0 & \\ [1ex]
$[56,2^+]_2$& $N(^2D_{S}){\frac 32}^+$ & & $-\frac 1{3\sqrt{15}}\frac {{\bf k}^2}{\alpha^2}
\left (1+\frac {\alpha^2}{2m_q^2}\right )$ & $-\frac {{\bf k}^2}
{12\sqrt{15}m_q^2}$ \\ [1ex]
& $N(^2D_{S}){\frac 52}^+$ & & $-\frac 1{3\sqrt{10}}\frac {{\bf k}^2}{\alpha^2}
\left (1-\frac {\alpha^2}{3m_q^2}\right )$ & $\frac {{\bf k}^2}
{9\sqrt{10}m_q^2}$ \\ [1ex]
& $\Delta(^4D_S){\frac 12}^+$ &  & $-\frac {5{\bf k}^2}{72\sqrt{15}m_q^2}$ & 
\\ [1ex]
& $\Delta(^4D_S){\frac 32}^+$ & & 0 & \\ [1ex]
& $\Delta(^4D_S){\frac 52}^+$ & & $\frac {5\sqrt{5}{\bf k}^2}
{216\sqrt{7}m_q^2}$ & \\ [1ex]
& $\Delta(^4D_S){\frac 72}^+$ & & $\frac {5{\bf k}^2}{36\sqrt{105}m_q^2}$ &
\\ [1ex]
$[70,0^+]_2$ & $N(^2S_{M^\prime}){\frac 12}^+$ & & 
$\frac 1{18}\frac {{\bf k}^2}{\alpha^2}$ & 
$-\frac 1{18}\frac {{\bf k}^2}{\alpha^2}$ \\ [1ex]
\hline
\end{tabular}
\end{table}

\newpage

\section*{\bf Figure captions}

\begin{itemize}
\begin{enumerate}
\item  The $Q^{2}$ dependence of the longitudinal amplitudes $S^{P}_{1/2}$ 
for the resonances $S_{11}(1535)$, $D_{13}(1520)$ and $F_{15}(1680)$.

\item The $Q^2$ dependence of the spin dependent sum rule of $g_1(x,Q^2)$
in the resonance region. The solid and dash lines represent the 
calculations with and without the quantity $\sigma_{TS}$.

\item The $Q^2$ dependence of the spin dependent sum rule of $g_2(x,Q^2)$
in the resonance region. 
The solid and dash lines represent the calculations with and without the 
quantity $\sigma_{TS}$.  The dot-dashed line comes from Ref. \cite{burk},
see text.

\item  The estimate of the sum rule $\int_0^1 g_{1}^{p}(x,Q^{2})dx$.
The nonresonant contribution comes from the result in Ref. \cite{zz93}.
The solid and dash lines correspond to the evaluations with and without
the quantity $\sigma_{TS}$.
\end{enumerate}
\end{itemize} 
\end{document}